\documentclass[conference]{IEEEtran}

\usepackage{color, colortbl}
\usepackage{multirow}
\usepackage{subfigure}
\usepackage{transparent}
\usepackage{balance}
\usepackage{url}
\usepackage{graphicx}
\usepackage{paralist}
\usepackage{amsmath, amssymb, amsthm, framed, graphics}
\usepackage{bytefield}
\usepackage{todonotes}
\usepackage{algorithm}
\usepackage{algpseudocode}
\usepackage[group-separator={,}]{siunitx}
\usepackage[space]{cite}

\definecolor{LightCyan}{rgb}{0.88, 1, 1}
\definecolor{LightRed}{rgb}{1, 0.88, 0.88}
\definecolor{LightGreen}{rgb}{0.88, 1, 1}
\definecolor{LightYellow}{rgb}{1, 1, 0.88}
\definecolor{LightOrange}{rgb}{1, 0.92, 0.77}

\newcommand{\ndnname}[1]{{\footnotesize\texttt{#1}}}
\newcommand{\Adv}{Adv}
\newcommand{\content}{{\em content}}
\newcommand{\interest}{{\em interest}}

\newcommand{\contentobject}{{\em content object}}
\newcommand{\cnack}{\mbox{\sf {\cal c}NACK}}
\newcommand{\fnack}{\mbox{\sf {\cal f}NACK}}
\newcommand{\cnacks}{\mbox{\sf {\cal c}NACK}s}
\newcommand{\fnacks}{\mbox{\sf {\cal f}NACK}s}
\newcommand{\ignore}[1]{}

\begin{document}
\title{To NACK or not to NACK? \\{\huge Negative Acknowledgments in Information-Centric Networking}}
\author{\IEEEauthorblockN{Alberto Compagno}
\IEEEauthorblockA{Sapienza University of Rome\\
compagno@di.uniroma1.it}
\and
\IEEEauthorblockN{Mauro Conti}
\IEEEauthorblockA{University of Padua\\
conti@math.unipd.it}
\and
\IEEEauthorblockN{Cesar Ghali}
\IEEEauthorblockA{University of California, Irvine\\
cghali@uci.edu}
\and
\IEEEauthorblockN{Gene Tsudik}
\IEEEauthorblockA{University of California, Irvine\\
gts@ics.uci.edu}}


\maketitle

\begin{abstract}
Information-Centric Networking (ICN) is an internetworking paradigm
that offers an alternative to the current IP\nobreakdash-based Internet architecture.
ICN's most distinguishing feature is its emphasis on information (content) 
instead of communication endpoints. One important open issue in ICN is
whether negative acknowledgments (NACKs) at the network layer are useful
for notifying downstream nodes about forwarding failures, or requests for
incorrect or non-existent information. In benign settings, NACKs are beneficial for 
ICN architectures, such as CCNx and NDN, since they flush state in routers and
notify consumers. In terms of security, NACKs seem useful as they can help
mitigating so-called Interest Flooding attacks. However,
as we show in this paper, network-layer NACKs also have some
unpleasant security implications. We consider several types of NACKs and
discuss their security design requirements and implications. We also demonstrate
that providing secure NACKs triggers the threat of producer-bound flooding
attacks. Although we discuss some potential countermeasures to these attacks,
the main conclusion of this paper is that network-layer NACKs are best avoided,
at least for security reasons.
\end{abstract}

\begin{IEEEkeywords}
information-centric networking, named-data networking, content-centric networking, 
negative acknowledgement, NACK, security considerations.
\end{IEEEkeywords}

\section{Introduction}
\label{sec:intro}
The original Internet design aimed to provide end-to-end 
connectivity, allowing users (numbering in tens of thousands) remote access to shared computing
resources. The number of Internet users has since grown tremendously, 
reaching over three billion. They use a wide variety of applications:
from email to dynamic web, to content distribution. This great shift in Internet usage highlighted
design limitations of the IP-based design and motivated research to explore new architectures.

Named-Data Networking (NDN) is one such new architecture~\cite{NDN}. It is one of the five Future Internet 
Architecture projects funded by the U.S. National Science Foundations (NSF)~\cite{NSF-FIA}. NDN is an instance of 
Information-Centric Networking (ICN)~\cite{ahlgren2012survey} that branched out of the Content-Centric 
Networking project (CCNx) at the Xerox Palo Alto Research Center (PARC)~\cite{jacobson2009networking, 
content-centric}. Despite recent differences in features, NDN and CCNx  share the same basic ICN 
vision. Instead of establishing communication between a source and a destination (via packets, as
in IP), in order to exchange data, NDN and CCNx directly address \content\ using unique human-readable  
names.  A consumer requests desired \content\ by issuing an \interest\ carrying the \content\ name. Then, 
the network is in charge of finding and returning requested \content.
Moreover, content follows (in reverse) the exact path of the preceding interest(s)  back to the 
consumer. In addition, routers keep state information for all received interests 
in their Pending Interest Tables (PITs), along with the corresponding interfaces on which they are received. 
When a router receives content, it uses information in a matching PIT entry to forward the content to the 
correct downstream router, towards the consumer.

To facilitate efficient content distribution, NDN and CCNx introduce in-network content caching. 
An entity (a router or a host) can satisfy an incoming interest if a copy of requested content 
is found in the local cache. Whenever an interest can be neither satisfied locally nor
forwarded, NDN and CCNx adopt different behaviors 
depending on the release version. 

Since NDN's initial release, producers simply drop interests they 
cannot satisfy, while router behavior in case of forwarding failures is unclear. 
Even though~\cite{yi2012adaptive} proposed network-layer NACKs for notifying downstream 
routers about forwarding failures, there is currently no support for this feature in NDN. 
However, recent discussions~\cite{nacksndnsim2015} indicate that network-layer NACKs might be
adopted by NDN. 

For its part, CCNx implemented network-layer NACKs until version 0.8.2. However, in 
the latest release, CCNx team announced that NACKs are considered a higher-layer functionality and 
are no longer implemented at the network layer~\cite{mosko2015ccnx}. We believe that all these oscillations 
(over time, and in different but related projects) represent a strong motivation for the analysis provided 
in this paper.

The use of NACKs, as an alternative to simply dropping unsatisfiable interests, has some advantages. 
First, consumers can rely on NACKs to quickly identify non-existing content, instead of waiting for issued
interests to time out. Second, NACKs allow consumers to differentiate between cases of non-existing content 
and packet (interest) loss. 
In the latter case, consumers have to wait until an issued interest expires before attempting 
retransmission. Third, the use of NACKs can help mitigate the effects of Interest Flooding (IF) attacks~\cite{gasti2013and}. 
In such attacks, adversaries flood routers with non-sensical (unsatisfiable) interests in order to exhaust 
their PITs. Once the PIT of a router gets full, it drops new incoming interests, resulting in a denial-of-service
for legitimate interests. Since, like content, a NACK traverses, in reverse, the path a corresponding interest, 
it causes routers to remove corresponding PIT entries and thus release valuable resources. Finally, NACKs 
can be a useful tool in notifying downstream routers that received interests cannot be forwarded further. 
Routers can thus quickly react and pursue alternative paths.

Despite aforementioned benefits, we show that network-layer NACKs in CCNx and NDN
have important and interesting security implications. In doing so, we differentiate between 
Forwarding-NACKs and Content-NACKs.
To the best of our knowledge, this paper represents the first attempt to  address security considerations 
for NACKs in the ICN context. The intended contributions of this paper are:
\begin{itemize}
	\item We assess benefits and identify scenarios justifying the use of network-layer NACKs.
	\item We discuss security requirements for implementing NACKs.
	\item We show that na\"ive security for NACKs can facilitate DoS attacks against producers. 
	\item We describe experiments that demonstrate effects of NACK-based DoS attacks.
\end{itemize}
As mentioned earlier, NDN and CCNx are research projects with the same goal of popularizing the ICN paradigm. 
Both NDN and CCNx are candidates for the next-generation Internet architecture. Even in the case they will never see wide adoption, 
their designs are likely to influence the Internet of the future. Therefore, we believe that this paper is 
both timely and important, since it studies, from a security  perspective, one of the key ICN features.

This paper is organized as follows. In Section~\ref{sec:overview} we present an overview of 
NDN and CCNx architectures. In Section \ref{sec:general-nack}, we identify two types of NACK 
messages usable in ICN, \cnacks\ and \fnacks, and we discuss their design requirements from a security perspective in 
Sections~\ref{sec:cnack} and \ref{sec:fnack}, respectively.
Section~\ref{sec:discussion} discusses some potential methods for preventing producer 
flooding attacks imposed by introducing secure NACKs. We present the related work 
in Section~\ref{sec:related_work}, and we conclude in Section~\ref{sec:conclusion}.

\section{Overview}
\label{sec:overview}
This section overviews NDN and CCNx. It can be skipped with no loss of continuity, given 
some familiarity with basic ICN concepts and terminology. 

\subsection{NDN}
Unlike IP, which emphasizes end-points of communication and their names/addresses,  
NDN~\cite{jacobson2009networking,NDN} focuses on content and makes it named, addressable
and routable at the network layer. A content name is composed of one or more variable-length 
components opaque to the network. Component boundaries are explicitly delimited by 
``\ndnname{/}'' in the usual path-like representation. For example, the name of a WSJ's 
news homepage content for May 1, 2015 
might be: \ndnname{/ndn/wsj/news/05-01-2015/index.htm}. Large content can be split into 
intuitively named segment, e.g., chapter 13 of Netflicks movie "Argo" could be named: 
\ndnname{/ndn/netflicks/movies/argo.mp4/ch13/}. 

NDN communication follows the general {\em pull} model, whereby content is delivered to consumers 
only upon (prior) explicit request, i.e., each content delivery is triggered by a request for that content.
There are two types of NDN packets: interest and content. A consumer 
requests content by issuing an \interest\ packet. An entity that can ``satisfy'' a given interest, 
i.e., has the requested content in its Content Store, returns it immediately.  If content $C$ with name 
$n$ is received by a router with no pending interest for that name, it is dropped as being unsolicited. 
Name matching in NDN is prefix-based. For example, an interest for \ndnname{/ndn/youtube/alice/video-749.avi} 
can be satisfied by content named \ndnname{/ndn/youtube/alice/video-749.avi/37}.\footnote{However, the 
reverse does not hold, by design.}
Note that the term \contentobject\ refers to a segment of a content, while
\content\ denotes the entire content before segmentation takes place.

NDN content objects include several fields. In this paper, we are only interested in the following 
four:
\begin{itemize}
\item \texttt{Name}: A sequence of name components followed by an implicit digest 
(hash) component of the content recomputed at every hop. This effectively provides each content 
with a unique name and guarantees a match when provided in an interest.
\item \texttt{Signature}: A public key signature, generated by the content producer, covering 
the entire object, including all explicit components of the name. The signature field also includes 
a reference (by name) to the public key needed to verify it.
\item \texttt{Freshness}: A producer-recommended time for the content objects to be cached.
\item \texttt{Type}: It specifies the content type, e.g., \texttt{DATA} or \texttt{KEY}.
\end{itemize}
An NDN interest message includes the name of requested content. In 
most cases, the last  component of a name (hash) is not present in interests, 
since NDN does not provide a means for consumers to learn content hashes
beforehand.

There are three types of NDN entities/roles:\footnote{A physical entity (a host, in today's 
parlance) can be both consumer and producer of content.}
\begin{itemize}
\item {\em Consumer} -- an entity that issues interest packets for content packets.
\item {\em Producer} -- an entity that produces and publishes (as well as signs) content. 
\item {\em Router} -- an entity that routes interest packets and forwards corresponding content 
packets. 
\end{itemize}
Each NDN entity (not just routers) maintains these three data structures~\cite{CCNxNodeImplementation}:
\begin{itemize}
\item {\em Content Store} (CS) -- cache used for content caching and retrieval. From here on, we use 
the terms {\em CS} and {\em cache} interchangeably. Recall that timeout of cached content is specified 
in the freshness field.
\item {\em Forwarding Interest Base} (FIB) -- table of name prefixes and corresponding outgoing 
interfaces. FIB is used to route interests.
\item {\em Pending Interest Table} (PIT) -- table of outstanding (``pending") interests and  
corresponding sets of interfaces from which interests arrive.
\end{itemize}
When a router receives an interest for a name $n$, and there are no pending interests for the same name 
in its PIT, it forwards the interest to the next hop(s), according to its FIB. For each forwarded 
interest, a router stores some amount of state information, including the name in the interest and 
the interface on which it arrived. However, if an interest for $n$ arrives while there is already an 
entry for the same content name in the PIT, the router collapses the present interest, 
storing only the interface on which it was received. When content is 
returned, the router forwards it out on all incoming-interest interfaces, and flushes the corresponding 
PIT entry. Since no additional information is needed to deliver content, interests do not carry 
any {\em source address}.

A router's cache size is determined by local resource availability. Each router unilaterally determines 
what content to cache and for how long. Upon receiving an interest, a router first checks its cache to 
see if it can satisfy this interest locally. Therefore, NDN lacks also any notion of {\em destination 
address} -- content can be served by any NDN entity. Producer-originated content signatures allow 
consumers to authenticate received content, regardless of the entity that serves this content.

\subsection{CCNx}
Both NDN and CCNx projects used to share the same codebase originally implemented by PARC. In August 
2013, the two projects separated. 
Both codebases still sharing the basic design features outlined above. However, in December 2013, 
PARC released the roadmap for the new codebase, CCNx 1.0~\cite{mosko2013ccnx}, increasing the 
differences between NDN and CCNx.

Until version 0.8.2, CCNx used to provide NACK support by design. A NACK message is a content 
object containing no data, but the name of the requested content, and a type with value \texttt{NACK}. 
Following CCNx (and NDN) specifications, all content objects must be signed~\cite{zhang2010named}. Therefore, 
all CCNx NACKs are signed by their producers. As mentioned above, recent CCNx 1.0 specifications 
removed NACK generation at the network layer. This is because they were (re)considered as a higher-layer 
functionality.

For the rest of this paper, we use the terms NACKs, NACK messages and NACK objects interchangeably to 
refer to content objects with type \texttt{NACK}.

\section{NACKs in General}
\label{sec:general-nack}
In communication protocols, there are usually two 
ways to confirm whether a packet (message or segment) has been received: 
acknowledgments (ACKs) or negative acknowledgments (NACKs). In ACK-based protocols, 
a receiver informs the sender about all successfully received packets. In NACK-based protocols, 
a receiver informs the sender whenever it believes that a received packet is unrecognized, non-sensical or 
corrupted~\cite{tanenbaum2010computer}.

In the next sections, we consider network-layer NACKs from a security perspective. In particular, we discuss two 
types of NACK messages that might make sense in ICNs: Content-NACKs (\cnacks) and 
Forwarding-NACKs (\fnacks). For each type, we 
present its benefits for network entities (consumers, producers and routers) and specify security 
requirements. Then, we show that -- even with these requirements met -- introducing \cnacks\  
has negative security implications for producers and routers, while \fnacks\ are generally 
beneficial.

\section{Content-NACKs}
\label{sec:cnack}
A \cnack\ is a packet generated by a producer at the network layer: it indicates that a 
content -- with the name reflected in a received interest -- does not exist, i.e., has not
been produced or published. A \cnack\ is realized as a special kind of a content object, of type \texttt{CNACK}. 
One intuitive analogy (though at a higher layer) is the well-known ``HTTP 404 not found'' 
message~\cite{fielding1999hypertext}.

\subsection{Benefits}
\cnacks\ offer several benefits. On the consumer side, they help applications to:
(1) distinguish between packet loss and {\em content not found}, and (2) reduce 
waiting time for consumers, i.e., inform consumers faster than interest timeouts. 
For routers and producers, \cnacks\ can reduce the effects of Interest Flooding (IF) attacks. 
Recall that a router creates a PIT entry for each distinct interest that it forwards.\footnote{Clearly,
this excludes collapsed interests.} A PIT entry is not purged until content arrives (from upstream),
gets cached and forwarded downstream. However, if an interest requests 
some non-existing content and the producer simply drops such interest, corresponding PIT entries
at all intervening routers (and at the consumer) remain until they expire. A producer-generated \cnack\
allows routers to purge PIT entries earlier and thus free their resources early. Even though this strategy 
does not fully mitigate the impact of IF attacks, it significantly reduces their effects.

In both NDN and CCNx, a router that receives a new interest (i.e., there is no PIT or cache hit)
might determine -- based on its local FIB -- that multiple outgoing interfaces are possible for forwarding.
If so, a router either: (1) forwards the interest on multiple interfaces, or (2) forwards the interest on 
one interface; in case of a time-out, it tries the next possible interface, and so on.\footnote{Other 
forwarding strategies are possible. However, we are not considering them here.} 
In the latter case (2), a router might incur considerable delay by sequentially trying (and timing out on)
every viable interface. However, recall that \cnacks\ are generated by the producer to indicate non-existing content. 
Therefore, if a router receives a genuine \cnack, trying other possible interfaces would be useless. This 
early detection of non-existent content is another advantage of \cnacks. Finally, since a \cnack\ is 
a type of content and is thus cached, subsequent interests for the same name are satisfied accordingly.

\subsection{Security Issues}
Despite aforementioned benefits, \cnacks' security implications 
should be carefully examined. We believe that support for insecure \cnacks\ 
opens the door for simple content-focused DoS attacks. 
Assume that an adversary \Adv\ controls a network link 
and can inject \cnacks\ for interests traversing that link. In this case, \Adv\ 
can prevent consumers from obtaining legitimate extant content. Even if there are 
multiple paths to the producer, \Adv\ only needs to inject \cnacks\
on just one path to succeed in the attack. For example, this attack can be trivially
exploited to enforce censorship over content considered subversive or
simply undesirable. 

More generally,  an unsecured \cnack\ -- being a special type of content -- can be abused
to essentially {\em poison} router caches. As described in \cite{gasti2013and}, a content
poisoning attack can occur in either reactive or proactive mode. The former corresponds
to the adversarial scenario above. The latter involves \Adv\ that, anticipating demand for
certain content, issues one or more bogus interests (perhaps from strategically placed zombie
consumers), ahead of genuine interests being issued. \Adv\ then replies with fake content
(from a set of compromised routers or compromised producers, at or near
genuine producers) thus pre-poisoning the caches of all routers that forwarded bogus interests.  

One variation of proactive content poisoning attack is even simpler. Again, predicting 
the name of content that has not yet been produced, \Adv\ issues an interest for such 
content and receives a legitimate \cnack\ from the genuine producer. Routers on the path
cache this \cnack. Even if the actual content is published soon thereafter, subsequent 
interests for that content will be satisfied with a cached \cnack, thus resulting in a DoS.

The above clearly motivates securing \cnacks, which intuitively 
translates into two requirements: (1) authenticating \cnack\ origin and integrity, i.e., detect fakes, 
and (2) checking \cnack\ freshness, i.e., to detect replays. We discuss these in the next section.

\subsection{Securing \cnacks}
Addressing authentication of \cnack\ origin and integrity is easy: in fact, one of the basic tenets 
of both NDN and CCNx is that all content
must be signed by its producer. (Indeed, creating a special type of content that is exempt from
being signed would violate this very tenet.) A router can elect to verify content signatures before caching 
or forwarding content. However, this process is not mandatory for several reasons 
discussed in \cite{ghali2014network}, hence triggering the aforementioned \cnack\ poisoning 
attack.\footnote{This is an attack similar to the content poisoning attack 
described in~\cite{ghali2014needle}.} For this reason, the Interest Key 
Binding (IKB) rule has been introduced in~\cite{ghali2014network}. Enforcing the IKB rule
requires consumers and producers to collaborate in order to provide routers with 
the trust context needed to verify only one signature per content. In particular, consumers include the digest of 
the producer's public key in every interest. For their part, 
a  producer includes its public key in all content it serves. A router that receives a content: 
(1) ensures that the digest of the public key in that content header matches the one provided in the matching interest, 
and (2) verifies the content signature with the public key included in the content header.\footnote{This 
assumes that fast public key cryptographic operations will be supported in
hardware in future routers.} The IKB rule would thus prevent \cnacks\ from being modified or generated by 
entities other than legitimate producers.

A complementary means of preventing content poisoning is via  
Self-Certifying Names (SCNs)~\cite{ghali2014network}. With SCNs, a consumer specifies, 
in the interest packet, the hash of the expected 
content. Using SCNs, routers only need to verify that the hash of a received content matches the value 
specified in the interest. A key advantage of this approach is that a content that is matched in this 
manner does not need to be signed. SCNs are particularly appropriate for static nested content, e.g., catalogs.

However, \cnacks\ cause a problem for routers when consumers use SCNs. Suppose that a benign consumer
requests a content using SCN in an interest. Even though a consumer might have pre-obtained the hash of currently 
requested content from a legitimate source (e.g., a catalog that it previously obtained using IKB), the content
in question could be no longer available from its producer, for various reasons. 
In that case, the producer would satisfy the interest with a \cnack. However, the hash of the latter 
would certainly not match the content hash reflected in the SCN from the interest. Therefore, 
such a \cnack\ would be dropped by routers as an invalid content. 
Fortunately, this problem can be solved by a minor modification to network-layer trust
rules proposed in~\cite{ghali2014network}: interests bearing SCNs should also (as a backup)
adhere to IKB, i.e., reflect the producer's public key, in order to handle (via signed \cnacks) 
expired or simply no-longer-available content.

Although the motivation for producer-signed \cnacks\ is not surprising, why this process 
should take place at run-time might not be obvious. First and foremost, producers cannot 
create and pre-sign \cnacks\ for all possible non-existing content. The reason is because 
content names can have arbitrary suffixes, resulting in an infinite number of possible names. 
In other words, a producer responsible for a name prefix \ndnname{/ndn/x/y/z}, should be ready
to respond to (in particular, by generating a signed NACK) an interest  requesting any content 
name starting with that prefix.

To prevent replay attacks, signed \cnacks\ must include a challenge by the consumer, and/or 
a timestamp set by producers. However, both means have certain drawbacks:
\begin{compactitem}
\item If each interest contains a unique consumer-selected challenge, then caching a signed \cnack\
that also includes this challenge is useless for other consumers who issue interests for the same content
at, or near, the same time.  Caching such a \cnack\ is beneficial only in the case of packet error or loss and 
retransmission. Moreover, PIT interest collapsing becomes a problem, since each interest to-be-collapsed
would have a different challenge. Thus, we conclude
that consumer challenges are problematic in the \cnack\ context. 
\item If, instead of challenges, each \cnack\ contains a producer-set timestamp, a time window 
needs to be defined to allow for transmission and caching delays. The selection of this window poses a 
problem. If too large, \cnack\ objects can be replayed for a longer time; else, if the 
window is too small, the probability of successful replay attacks decreases, while the probability of 
\cnacks\ wrongly considered invalid increases. 
One viable alternative is to use producer-specified expirations for signed and time-stamped \cnacks.
This would address \cnack\ replay attacks. Nonetheless, we note that timestamps require a global 
synchronization protocols, e.g., a secure version of NTP \cite{mills2010rfc}.
\end{compactitem}
\ignore{
As mentioned above, \cnacks\ are treated similar to content objects with type \texttt{DATA}, thus, they 
might be cached by routers. Although challenge-based \cnacks\ only benefit from the cache in case of 
retransmission, timestamp-based \cnacks\ can be cached and used to satisfy future interests. However, there 
are some considerations regarding the latter's \texttt{Freshness}. Producers must not set \texttt{Freshness} 
to be larger than the accepted time window. If a cached \cnack\ is served after the window elapsed, it will 
be considered invalid and dropped by receiving consumers. Moreover, \cnacks\ are cache-agnostic. Imagine an 
network architecture that follows the exact same design and requirements as NDN/CCNx but does not implement 
in-network caching. In this case, secure \cnacks\ can still be implemented. The only difference is that all 
interests requesting non-existing content must reach their corresponding producers in order 
to be satisfied with \cnacks. Furthermore, if this hypothetical architecture dictates routers to include 
traversed path information in interests instead of using PITs, interest flooding attacks on routers cannot 
be mounted as explained above. Therefore, \cnacks\ will no longer be helpful in reducing the effects of such 
attacks, but can still be beneficial for informing consumers about requested non-existing content.
} 
Based on the above discussion we summarize the requirements for securing \cnacks:
\begin{compactenum}
\item Signature: a \cnack\ must be signed by its producer, just like any other content.
\item Timestamps: a \cnack\ should be generated, not per interest, but per time interval.
\item Expiration: a \cnack\ for plausible content (e.g., not yet published) should include expiration time.
\end{compactenum}

\subsection{Secure \cnacks: a Blessing or a Curse?}\label{subsec:secure-cnack_implications}
Unfortunately, secure \cnacks\ that satisfy our three requirements (which are themselves 
motivated in part by DoS prevention) facilitate producer-focused DoS attacks.
Such an attack occurs when \Adv\ sends a large number of closely-spaced interests requesting 
non-existing (and possibly non-sensical) content. A producer that receives a barrage of these
interests generates a \cnack\ for each one, which requires generating a signature. The resultant
computational load on the producer could be overwhelming. Furthermore, large numbers of
useless \cnacks\ would pollute router caches.

Note that generating one \cnack\ for all interests arriving within a certain time interval is not effective
against this DoS attack. This is because a smart \Adv\ -- instead of issuing interests for the same (non-existent)
name -- would issue many interests, each for a distinct name composed of a common prefix (registered to 
the victim producer) and a random suffix, e.g., \ndnname{/ndn/cnn/news/world/\$\&F(?\%}. 
One simple countermeasure is to allow producers to issue \cnacks\ for prefixes. For example, a \cnack\
for \ndnname{/ndn/cnn/news/world/}, once cached in routers, would throttle all interests with that 
prefix, including non-sensical ones. However, the very same \cnack\ would result in DoS for legitimate 
interests, e.g., referring to \ndnname{/ndn/cnn/news/world/china}.

The discussion above leads us to a logical conclusion that secure \cnacks\ should be implemented carefully.
Specifically, a producer must first decide whether an incoming interest is plausible or non-sensical.
An interest is {\em plausible} if the producer believes that the referenced content name 
might have existed in the past or might exist in the future. In contrast, an interest is {\em non-sensical}
if it refers to implausible (or unlikely to ever exist) content name. 
We have no guaranteed way of distinguishing between these two types of interests. This task
is perhaps best left up to individual applications. 
As far as producer's strategy, we believe that a producer should have the option of replying 
with a secure \cnack\ in response to a plausible interest. Otherwise, a producer should not
reply at all to a non-sensical interest. This prompts the addition of another requirement for securing
\cnacks:
\begin{compactenum}
\item[4)] Plausibility: a \cnack\ should be generated only for a plausible interest.
\end{compactenum}

\subsection{Experimenting with Secure \cnacks}
To assess the efficacy of producer-focused DoS attacks, we performed several experiments, 
using ndnSIM~\cite{afanasyev2012ndnsim,ndnSIM}, to demonstrate additional overhead imposed by 
generating a network-layer \cnack\ per interest. Although, as discussed above, secure \cnacks\ 
should be generated only for plausibly named content, 
a smart \Adv\ can still generate many names that a producer application can consider to be plausible. 
This can be caused by poorly implemented applications, or by the difficulty of distinguishing 
plausible from non-sensical names.

\begin{figure}[t]
\fbox{\centering{
	\includegraphics[width=0.98\columnwidth]{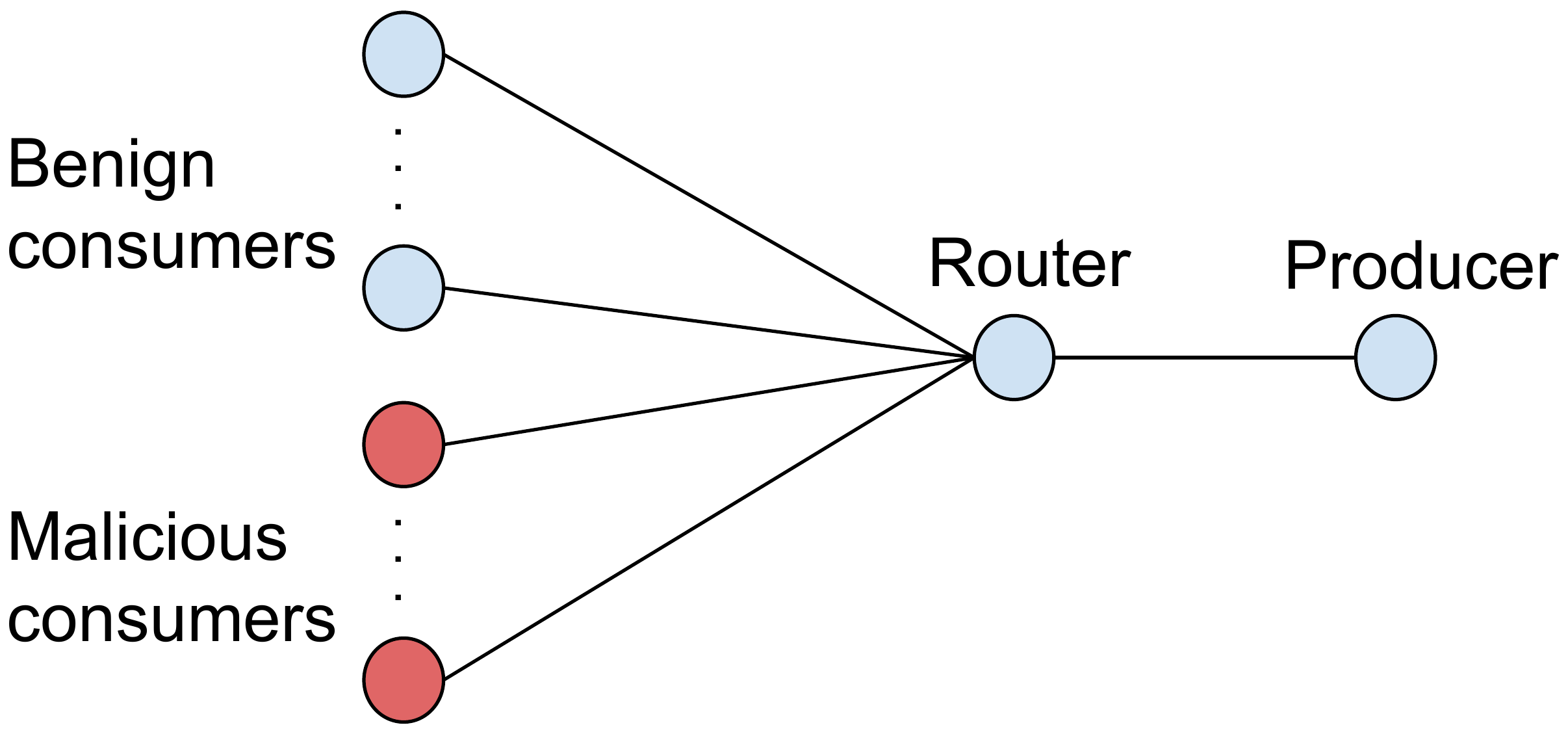}}}
	\caption{Simulation Topology.}
	\label{fig:scenario}
\end{figure}

In our experiments, we consider the simple network topology illustrated in Figure 
\ref{fig:scenario}. Also, we let benign and malicious consumers issue a large number of interests to a single producer 
at different rates: benign consumers send $10$ interests per second for existing content; while malicious 
consumers send $100$ non-sensical interests per second. We implemented two consumer modes:
\begin{compactenum}
\item Basic: consumers request sequential content under a specific name space, e.g.,
\ndnname{/ndn/a/1}, \ndnname{/ndn/a/2}, etc.
\item Advanced: content requested by consumers adheres to a Zipf distribution. This reflects practical 
applications where some content is more popular than other.
\end{compactenum}
%

\begin{figure}[bht]
\centering
\includegraphics[width=\columnwidth]{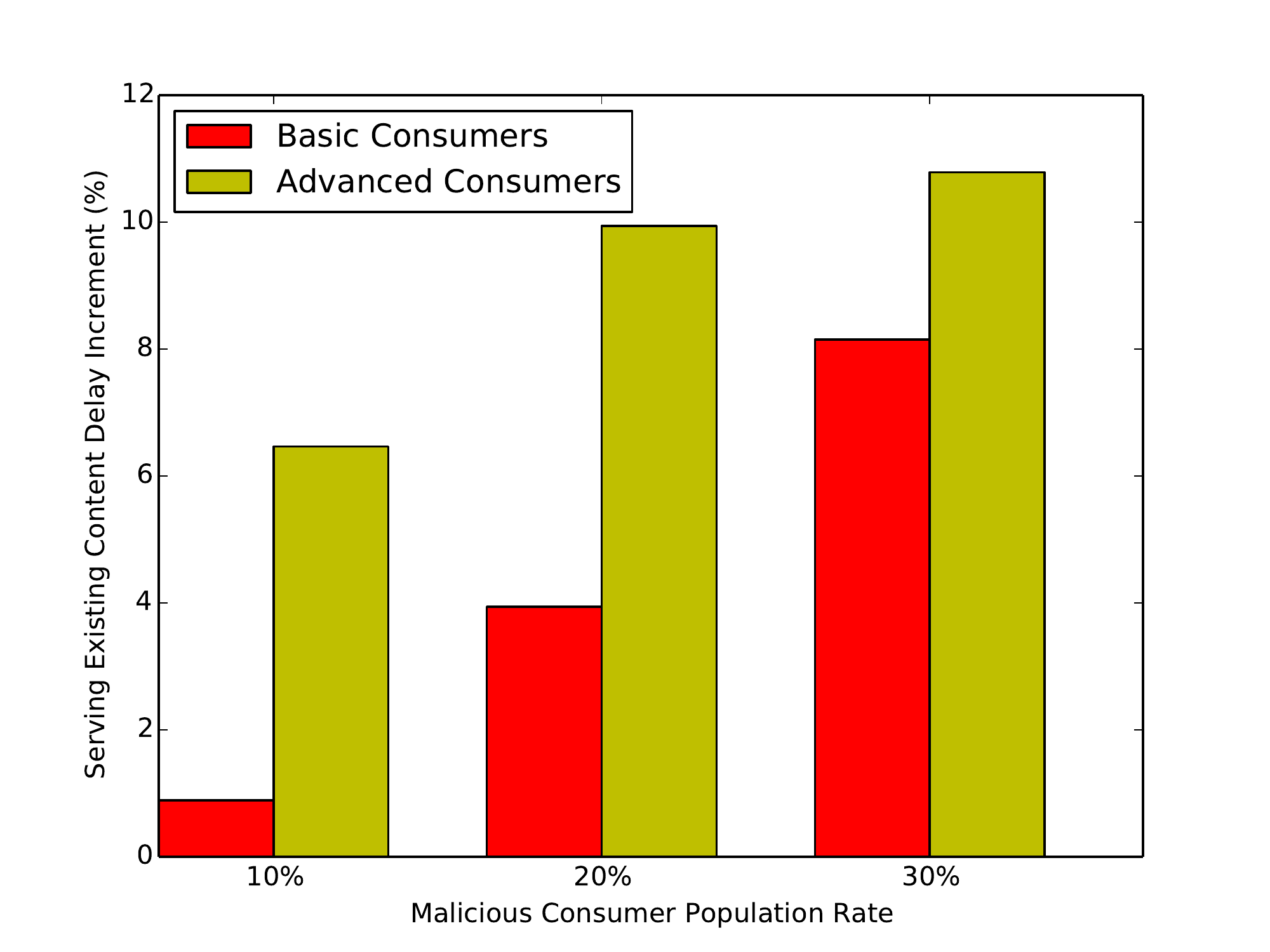}
\caption{Serving existing content delay increment compared to the base case for varying malicious 
consumer population rates.}
\label{fig:average_time}
\end{figure}

Figure \ref{fig:average_time} shows the delay increment in serving existing content, for both basic and advanced 
benign consumers. In the base case all consumers are benign. The results show 
the additional time required by the producer to serve existing content, as compared to the base case, for different 
malicious consumers population (MCP) rates (10\%, 20\%, and 30\%). As expected, increasing the number of malicious consumers, 
increases the producer overhead when serving existing content. Moreover, this overhead increases when using advanced 
consumers. 
This behavior is motivated as follows: collapsing of interests requesting existing content reduces the number of these interests on the link between the router and the producer. In fact, reducing interests on this link allows the router to forward more non-sensical interests to the producer. Therefore, the latter is forced to generate and sign more \cnacks.

\begin{figure}[t]
\includegraphics[width=\columnwidth]{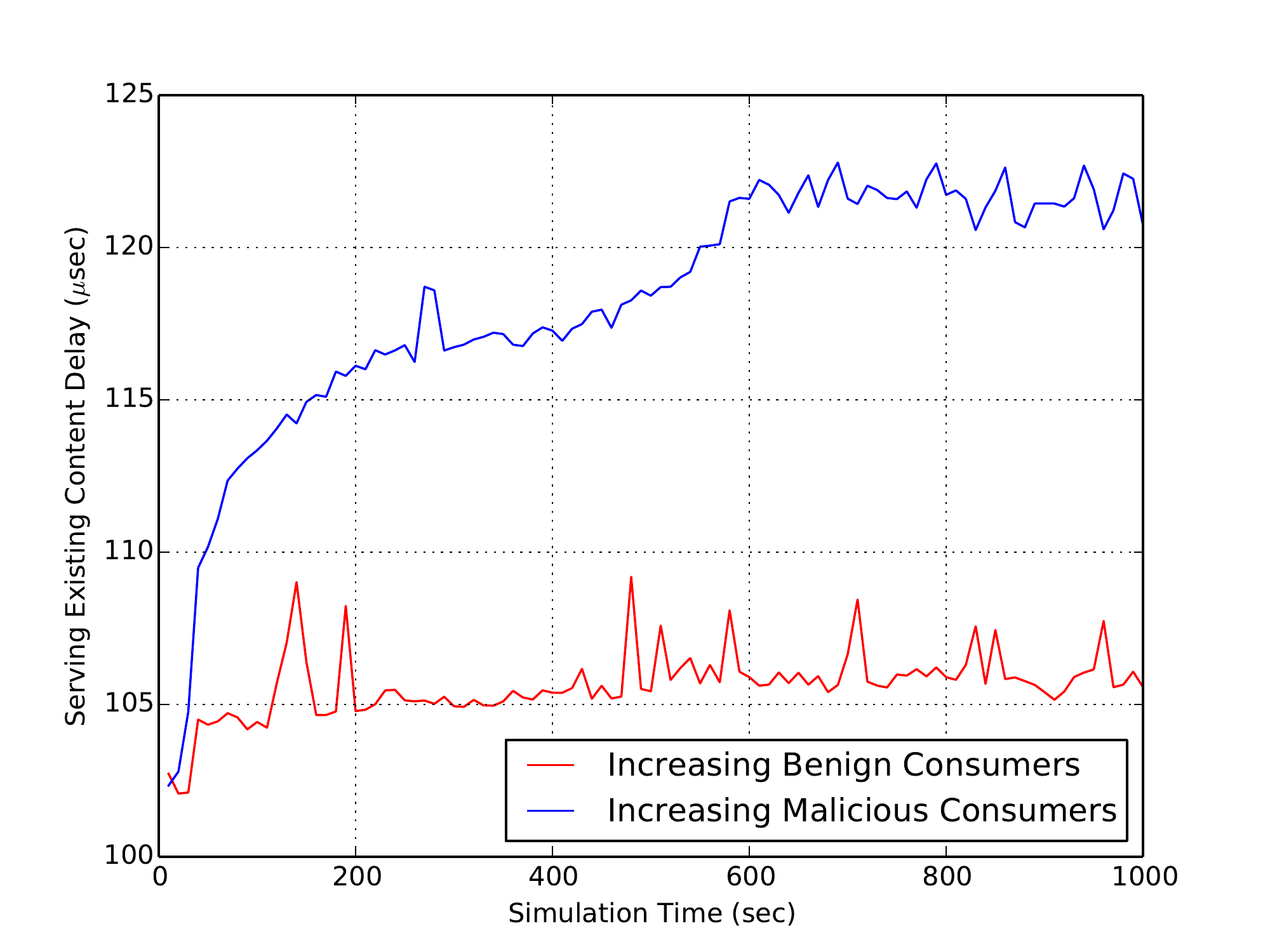}
\caption{Serving existing content delay for gradually increasing number of consumers.}
\label{fig:increasing_time}
\end{figure}

We also studied experimentally the delay in serving existing content, when the number of 
consumers increases. In particular, we started the simulation with 200 benign consumers, and we considered two scenarios: (1) adding one benign consumer per second; (2) adding a malicious consumer per second. In both cases, we stop adding nodes after 500 seconds, and measure the delay in serving content until the 1000-th second of simulation.
The result of this experiment is illustrated in Figure \ref{fig:increasing_time}.
We note that increasing 
number of benign consumers does not significantly affect the producer performance, while increasing the 
number of malicious consumers does (e.g., after 500 seconds, the delay is some 10\% more than the case with only benign nodes).

\section{Forwarding-NACKs}
\label{sec:fnack}
A \fnack\ is a packet generated by a router at the network layer. Its purpose is to inform downstream 
routers that an interest cannot be forwarded due to congestion or unknown next hop~\cite{yi2012adaptive}.
Since edge routers are usually configured with a default route to an upstream router, \fnacks\ generated 
due to unknown next hop are most likely to occur at the network core. A good analogy to \fnacks\ is ICMP 
destination unreachable message~\cite{postel1981rfc}.

Recall that, in both NDN and CCNx, a router's FIB might specify multiple interfaces on which an interest 
with a particular name (prefix) can be forwarded. In such cases, a router has two forwarding choices: 
\begin{enumerate}
\item { {Parallel:}} Forward the interest on all specified interfaces at the same time and set 
either the same or various time-outs for each interface. The PIT entry is flushed if all interfaces time out.
\item { {Sequential:}} Forward the interest on one interface and wait; in the event of a time-out, 
try another interface, and so on. Once the last possible outgoing interface times out, the PIT entry is flushed.
\end{enumerate}
We distinguish the cases of a router {\em generating} and {\em forwarding} \fnacks.
There are two reasons for a router to generate an \fnack: (1) FIB lookup failure, i.e., an entry indicating 
the next-hop of the received interest does not exist, or (2) all FIB-specified outgoing interfaces are congested.
A router that generates an \fnack, sends it out on each interest incoming interface listed in the appropriate
PIT entry. It then flushes the PIT entry.

\begin{figure*}[t]
\centering
\subfigure[\fnack\ parallel interest forwarding strategy.]
{
	\includegraphics[width=\columnwidth]{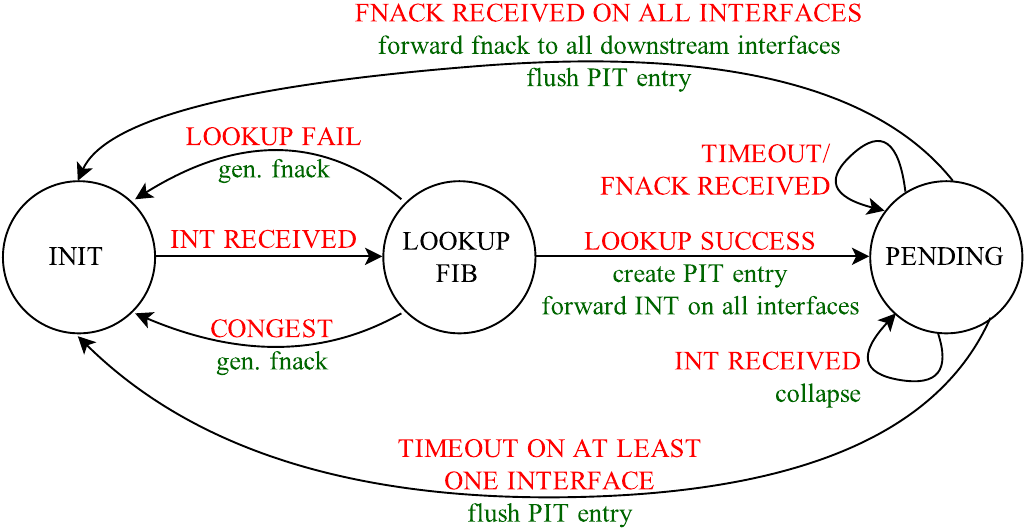}
	\label{fig:fnack_parallel}
}
\subfigure[\fnack\ sequential interest forwarding strategy.]
{
	\includegraphics[width=\columnwidth]{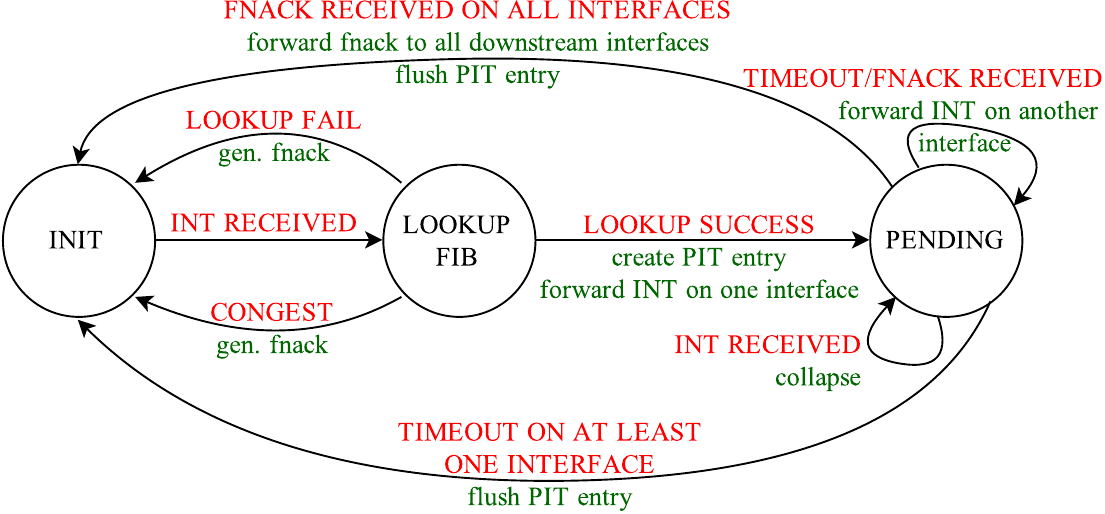}
	\label{fig:fnack_sequential}
}
\caption{\fnack\ generation and forwarding state diagrams (red/upper case: events, green/lower case: actions.)}
\label{fig:fnack_state_diagrams}
\end{figure*}

A router must forward \fnacks\ on all downstream interfaces (on which interests were received)
if it receives an \fnack\ on {\em every} upstream interface specified in the FIB, regardless of 
whether parallel or in sequential forwarding is used. Conversely,
if an \fnack\ is not received on at least one upstream interface (i.e., at least one time-out occurs)
a router {\em must not} forward \fnacks\ downstream. This is because 
a time-out does not imply producer unreachability. A producer 
might have actually received the interest and decided to drop or ignore it. 
Figure \ref{fig:fnack_state_diagrams} shows two state diagrams (one for parallel and the other -- for sequential case) 
for generating and forwarding \fnacks.

\subsection{Securing \fnacks}
Similar to \cnacks, insecure \fnacks\ trigger content-focused DoS attacks. \Adv\
controlling a link can inject fake \fnacks\ in response to interests on that link. This would 
prevent consumers from obtaining requested content.

Securing \fnacks\ seems similar to doing the same for \cnacks, i.e., ideally we would need origin 
authentication and replay prevention. However, we cannot use the methods from Section~\ref{sec:cnack}. 
If we require each \fnack\ to be signed, \Adv\ can easily generate many spurious interests  
that cannot be forwarded by a particular router. That victim router would then be forced to sign one
\fnack\ for each spurious interest. Since signing is often appreciably more expensive than verification (e.g., in RSA),
computational overhead for the victim router would easily translate into a full-blown DoS attack.\footnote{Note
that some digital signature techniques flip this balance, e.g., in DSA, verification is more expensive than
signing. However, the DoS attack would then be even worse, since multiple routers would verify \fnack\
signatures.}

Furthermore, \fnack\ signing would trigger the need for a routing PKI since verifying \fnack\ signatures
cannot be done mechanically: public key certificates must be fetched, verified and revocation-checked.
This represents another challenge for supporting signed \fnacks.

\ignore{
Mechanically verifying \fnack\ signatures does not determine whether they are generated by benign upstream 
routers or \Adv\ controlling upstream links. This opens the door for \fnack\ forgery and poisoning 
attacks. Moreover, signature verification can be abused to mount DoS attacks against 
downstream routers. \Adv\ can trigger $R$ to generate signed \fnacks\ and send them to its downstream 
neighboring router $R'$. This causes the latter to verify each \fnack\ signature, which requires fetching 
and verifying corresponding public keys. The affect of these DoS attacks are more devastating than their 
counterpart against producers because they hinder routers ability to process incoming packets, paralyzing 
segments of the network.
}

\ignore{
We note that the IKB rule cannot facilitate signature verification in \fnacks. Recall that, in IKB, 
a consumer must specify the public key of the {\em producer} that is expected to sign the content. If IBK 
is applied, downstream routers will not accept \fnacks\ as a legitimate response. This is because their verifying 
keys do not match the ones specified in the corresponding interests. To solve this problem, consumers need 
to specify the public keys of all routers on the path to the producer. In this case, routers receiving 
\fnacks\ should ensure that its public key digest is specified (among several) in the corresponding 
interest. This, however, (1) requires a path discovery-like protocol, (2) increases the size of interests, 
and complexity of interest collapsing, and (3) more importantly, increases routers overhead.
}

However, if we assume that trust relationships can be established between neighboring routers, 
\fnack\ authentication can be easily achieved. In this case, \fnacks\ can be sent downstream over 
a sequence of pair-wise secure channels between neighboring routers. We can safely assume
that such long-term channels are maintained between every pair of adjacent NDN or CCNx routers.
One trivial way of securing \fnacks\ hop-by-hop is by using a keyed hash, HMAC~\cite{bellare1996keying}.
Replay prevention can be achieved via timestamps, especially considering that adjacent routers are
likely to maintain closely synchronized clocks.

\subsection{Experimenting with Secure \fnacks}
We ran several experiments using ndnSIM 2.0 to demonstrate the negligible impact of secure \fnacks. 
We used the same topology as in Figure \ref{fig:scenario}. Benign consumers request 10 
contents per second, while malicious consumers send 100 interests per second; these interests 
cannot be forwarded by the router. Our evaluation metric is processing time for the router to forward an 
interest towards the producer. All consumers (benign and malicious) implement the basic mode.

\begin{figure}[t]
\centering
\includegraphics[width=\columnwidth]{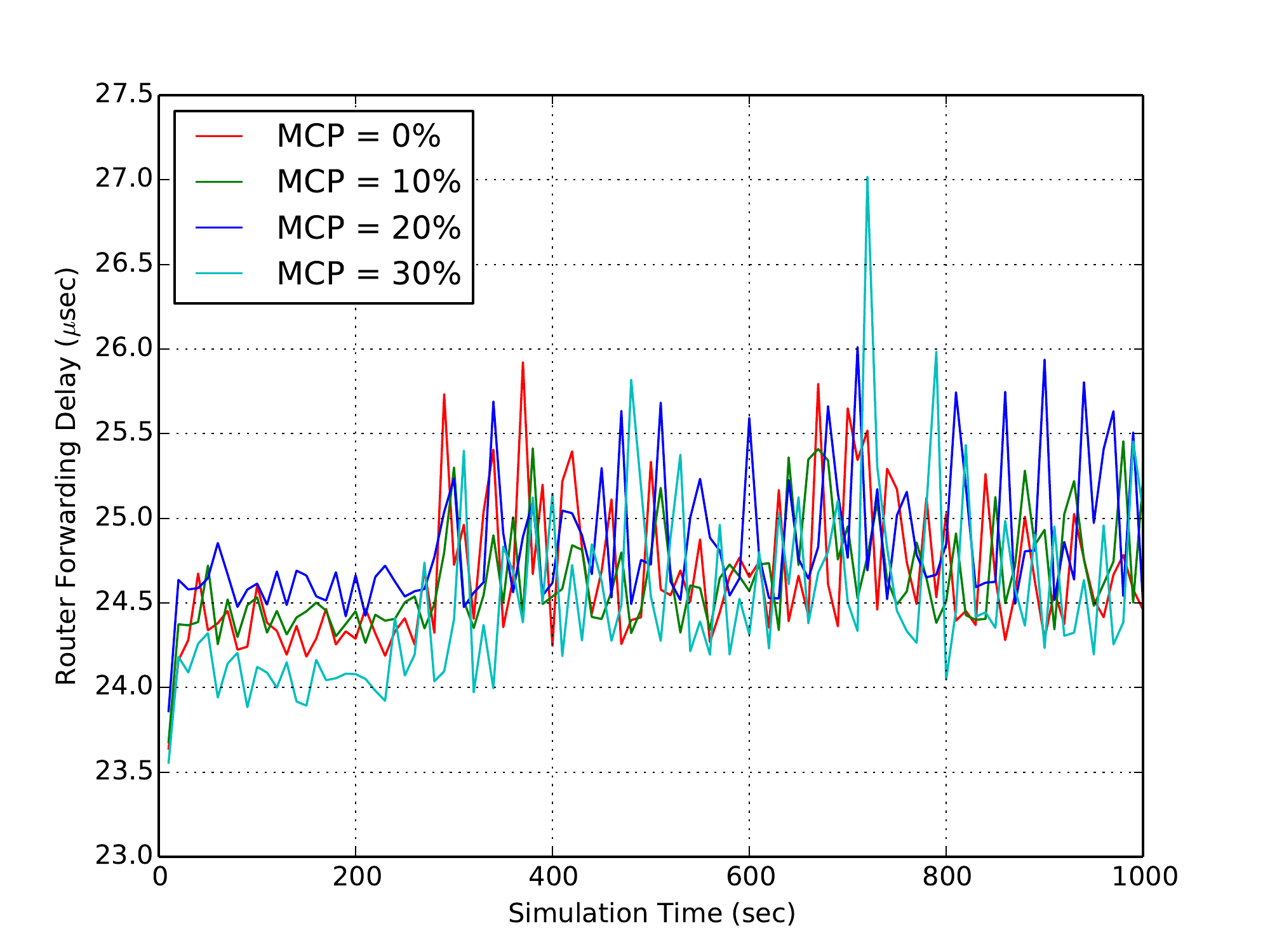}
\caption{Router forwarding performance for different MCP rates.}
\label{fig:fncak_diff_mcp}
\end{figure}

\begin{figure}[t]
\centering
\includegraphics[width=\columnwidth]{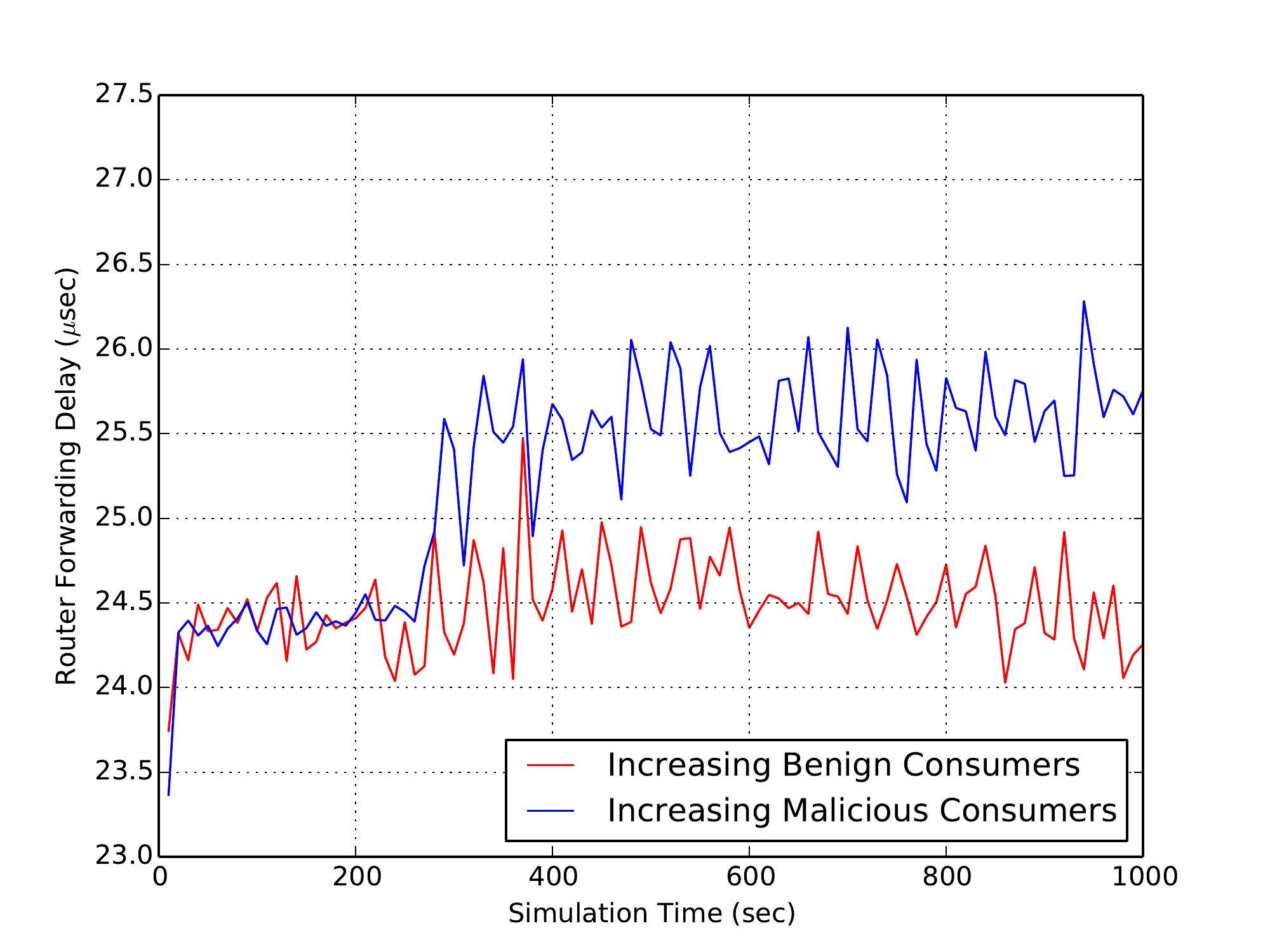}
\caption{Router forwarding delay for gradually increasing number of consumers.}
\label{fig:increasing_fncak}
\end{figure}

We implemented two scenarios. In the first, we compared router forwarding performance 
for different rates of MCP (0\%, 10\%, 20\% and 30\%). The total number of consumers in this scenario is 200. 
Figure \ref{fig:fncak_diff_mcp} shows that even with 30\% MCP rate, router forwarding performance is not
affected. In the second scenario, the number of consumers increases gradually. Initially, there are 200 benign 
consumers. We then either: (1) increase the number of benign consumers (one every second) until reaching 700, 
or (2) introduce 500 malicious consumers (one every second). Figure \ref{fig:increasing_fncak} illustrates 
the results: for up to 300 malicious consumers, router performance  
is similar to the case where the network contains only benign consumers. Even if the number of malicious 
consumers exceeds 300, router performance decreases only by an average of 4\%.

\section{Mitigating Producer-Focused DoS Attacks}
\label{sec:discussion}
As discussed earlier, securing \cnacks\ comes at a price of possible DoS attacks on content producers. 
We now discuss some ways to mitigate the impact of such attacks.

One approach is to separate content-serving and \cnack-generation activities. Producers can 
set up special-purpose gateways that distinguish between interests requesting existing and non-existing content. 
The former are forwarded to the actual content repository that serves requested content, while the latter are 
forwarded to a special server that generates and signs \cnacks. However, this only works for 
static content because producers need to keep gateways updated with all published content, 
which cannot be achieved for (dynamic) content generated upon request.

By redirecting the attack towards the \cnack\ generation server, producers can still continuously 
serve content. However, the network still needs to deal with the attack traffic, which might consume 
a lot of bandwidth. Moreover, routers would have to create PIT entries for all  interests 
since they cannot differentiate between interests requesting existing and non-existing content. 
If routers were capable of such differentiation, DoS attacks would be preventable closer to their sources. 

One way of achieving this, is by allowing a producer to relay the list of all its published content names to routers.
A producer can use these names to construct a Bloom filter~\cite{bloom1970space} and 
disseminate it to routers processing interests for this producer. The dissemination of such filters 
depends on producer's policies. For instance, a producer can fall back on Bloom filters when its load of generating 
and signing \cnacks\ reaches a certain threshold. 

\begin{figure*}[t]
\centering
\subfigure[Variable filter size, number of elements in the set $\mathbb{S}$, and 
number of hash functions used.]
{
	\includegraphics[width=\columnwidth]{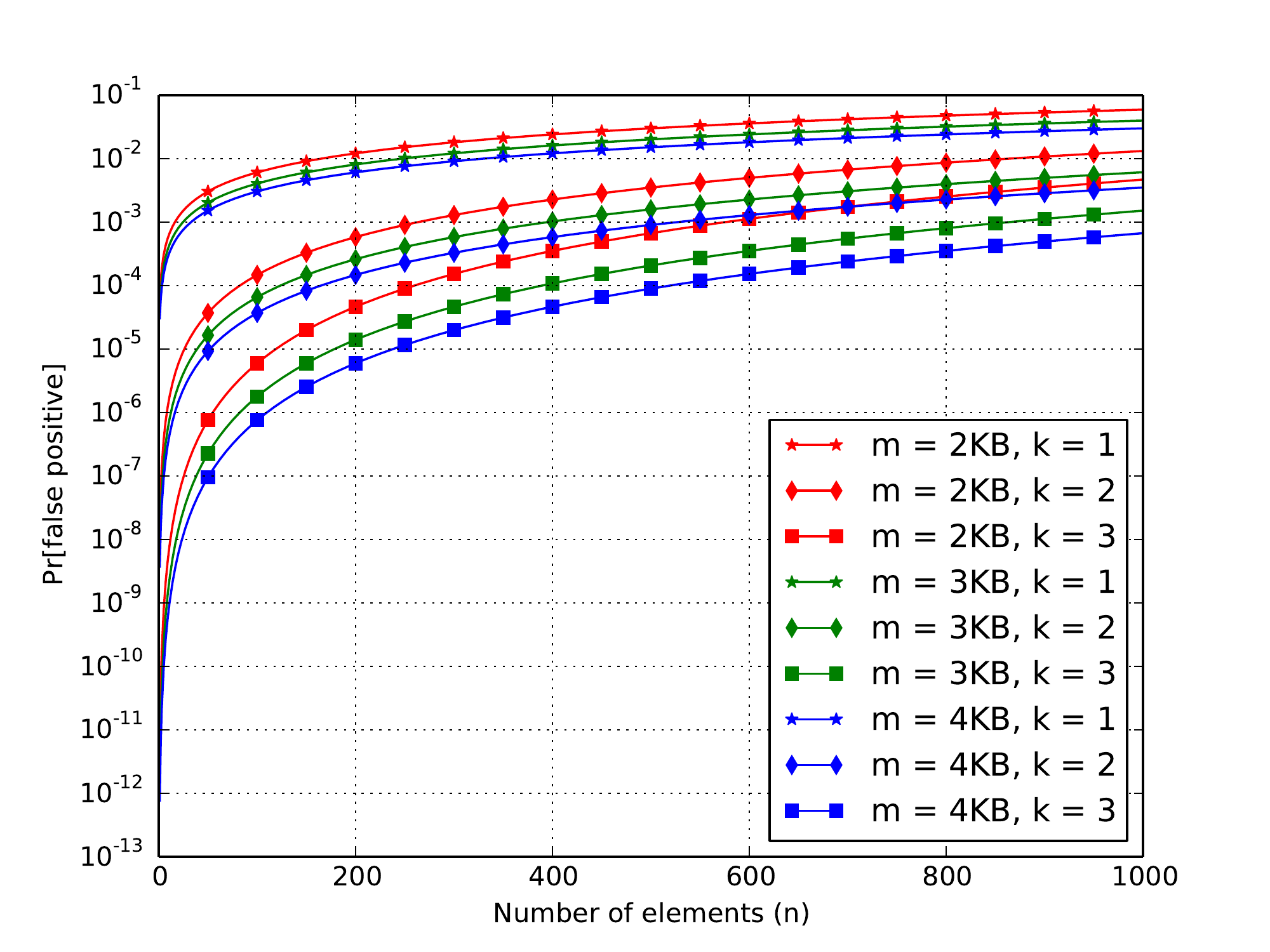}
	\label{fig:bloom_fpp_nopt}
}
\subfigure[Variable filter size and number of elements in the set $\mathbb{S}$, 
and optimized number of hash functions used.]
{
	\includegraphics[width=\columnwidth]{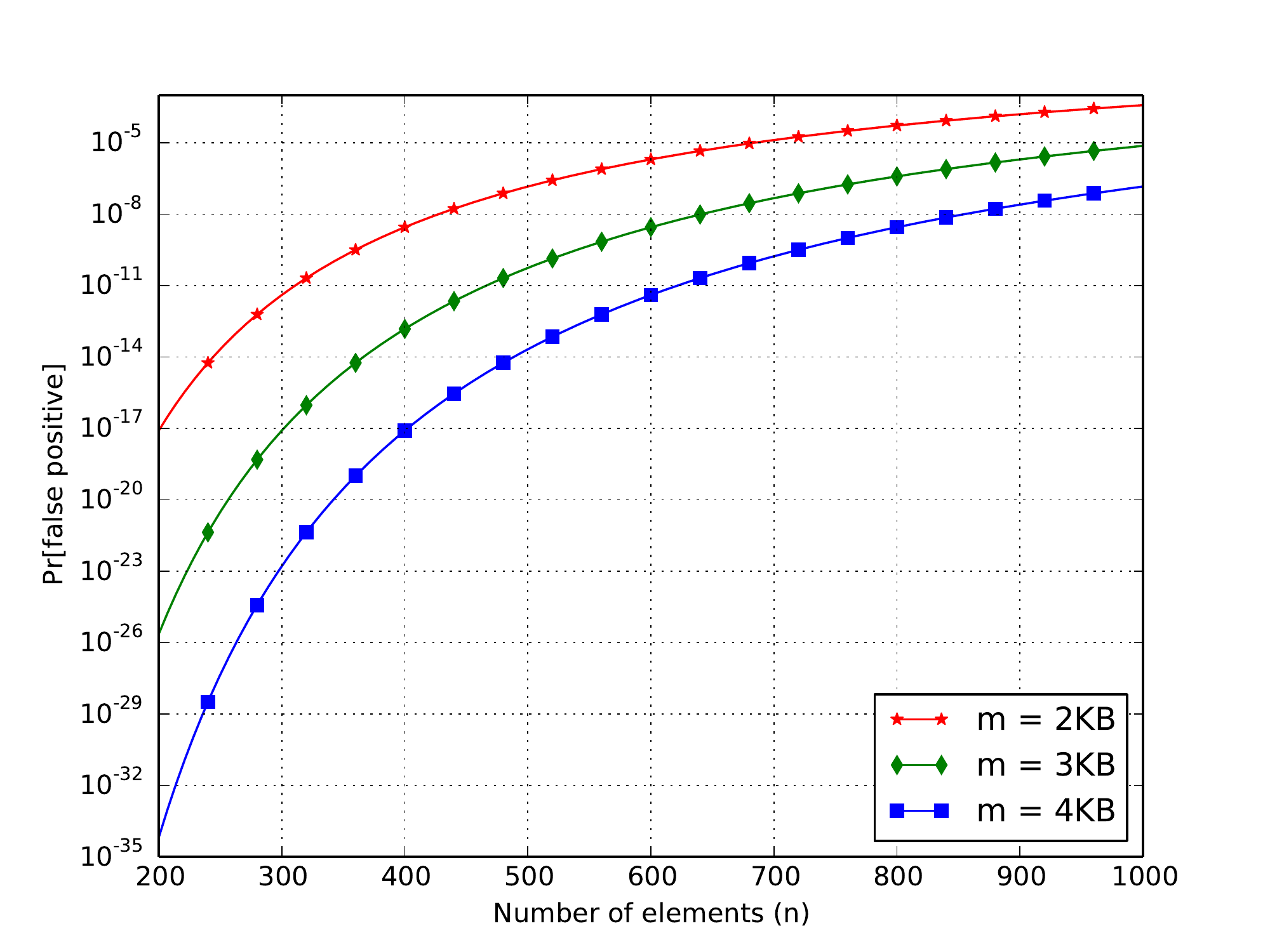}
	\label{fig:bloom_fpp_opt}
}
\caption{Bloom filter false positive probability.}
\label{fig:bloom_fpp}
\end{figure*}

Bloom filters are created by producers periodically, or whenever new content objects are published. These filters 
can be implemented as content objects with a specific type, e.g., \texttt{BLM-FLTR}. Therefore, they will be 
cached by routers and used to satisfy pending interests, thus clearing corresponding PIT entries. The only difference 
is that Bloom filters, if cached, will not be used to satisfy future interests. Moreover, the caching duration of 
these filters depends on the \texttt{Freshness} value included in their headers. Producers need to carefully set 
this value to be compatible with the frequency at which new content is being published. For instance, 
\begin{align}
\texttt{Freshness} = \frac{1}{\mathrm{avg}(f)|_{\tau}},
\end{align}
where the denominator represents the average value of content publishing frequency over a specific period of time 
$\tau$.

Furthermore, the size of a Bloom filter depends on the number of elements (content names) loaded into it. Recall 
that large content objects can be divided into smaller segments, each having a unique name. We claim that the size of 
each Bloom filter should be upper bounded by the maximum size of a content segment. This avoids the case where a 
Bloom filter is split into multiple segments, thus requiring multiple interests to request the whole filter. Since 
producers disseminate Bloom filters as a reply to a single interest, they should fit in a single content segment.

On the other hand, Bloom filter's false positive probability (illustrated in Figure \ref{fig:bloom_fpp}) depends 
on its size $m$ (in bits), the number of 
elements $n$ in the set $\mathbb{S}$, which are loaded into the filter\footnote{In this case, $\mathbb{S}$ is the set of published content names.}, and the number of used hash functions $k$. This false positive probability increases as $n$ and 
$k$ increase, and decreases as $m$ increases. Assuming that the hash functions used $(h_1, \dots, h_k)$ map each 
element of $\mathbb{S}$ into a random value uniformly distributed over the range $[1, \dots, m]$, the false 
positive probability can be expressed as in Equation~\ref{equ:bloom_fpp}~\cite{broder2004network}.
\begin{align}
\label{equ:bloom_fpp}
\Pr \left[ \mathrm{false\ positive} \right] &= \left( 1 - \left( 1 - \frac{1}{m} \right)^{kn} \right)^k \nonumber\\
&\approx \left( 1 - \mathrm{e}^{-\frac{kn}{m}} \right)^k
\end{align}

Figure~\ref{fig:bloom_fpp_nopt} illustrates the Bloom filter's false positive probability when varying $m$, $n$, 
and $k$. However, for a given $m$ and $n$, the number of hash functions $k$ can be optimized. In this case, the false 
positive probability can be calculated using Equation~\ref{equ:bloom_fpp_opt}~\cite{broder2004network}.
\begin{align}
\label{equ:bloom_fpp_opt}
\Pr \left[ \mathrm{false\ positive} \right] = \left( 0.6185 \right)^\frac{m}{n}
\end{align}

In practice, producers can optimizes the number of hash function in order to achieve lower false positive probability. 
However, an upper bound of $k$ can be set to limit the hashing overhead required by routers. Figure~\ref{fig:bloom_fpp_opt} 
demonstrates the Bloom filter's false positive probability when varying $m$ and $n$, and optimizing $k$.

Based on the plots in Figure~\ref{fig:bloom_fpp}, loading all published content names into a single Bloom filter (which size is upper bounded by the maximum size of a single content fragment), might not lead to a desired false 
positive probability. In this case, producers can create a separate Bloom filter for each namespace (or sub-namespace) 
they publish. Therefore, achieving the desired false positive probability might entail an upper bound on the number of 
content published under each namespace. It might also require redesigning the namespace hierarchy of producers 
implementing the aforementioned countermeasure. However, we will not discuss this optimization problem any further 
since we believe it is out of the scope of this paper. Moreover, the number of interests requesting non-existing content 
that forwarded to producers due to the probabilistic nature of Bloom filters, can be dramatically minimized with proper 
configuration of filters parameters.

Although Bloom filters are content objects that follow the same path, in reverse, of their corresponding interests, 
it is worth mentioning that they should not be delivered to consumers. The reason is because malicious consumers gain 
attack advantages when possessing these filters. For instance, \Adv\ can pre-compute a list of content names that pass 
verification and use it to launch a distributed DoS attack against the target producer.\footnote{We do not 
consider the case where malicious routers deliver Bloom filters to malicious consumers using side channels.} Moreover, 
such attacks can be circumvented if edge ISPs do not forward Bloom filters towards their customers, or filters are 
re-created periodically with different parameters.

\section{Related Work}
\label{sec:related_work}
In the current Internet, negative acknowledgments are proposed as a form of error notifications in error 
control methods. For instance, at transport layers, Automatic-Repeat-Request (ARQ) implements error control 
method in Go-Back-N and Selective Repeat~\cite{lin1984automatic}. In Go-Back-N, receivers detecting a packet 
loss send a NACK packet to the sender indicating the missing packet. In this case, the latter will restart the 
transmission from the lost packet. On the other hand, in Selective Repeat, receivers still use a NACK to notify 
a packet loss and the sender only resend that specific packet. Compared to Go-Back-N, Selective Repeat reduces 
the number of retransmissions. 

In broadcast (one-to-many) communications, NACKs are preferred over ACKs to reduce network congestion and 
packets collision \cite{pingali1994comparison}. 
The reason is because using selective NACKs allows reducing the number of packets sent by receivers, hence 
reducing the probability of packet collision. However, NACK based mechanisms are prone to NACK implosion. 
In case of packet loss, the sender receives many NACKs from all receivers. Stran et al. \cite{satran2004nack} 
propose a time-based mechanism to reduce NACK implosion. Every receiver detecting a packet loss initiates a 
random timer. The receiver having the shortest random interval unicasts a NACK to the sender, which immediately 
multicasts the NACK to the other receivers. All other receivers having the same missing packet thereupon 
suppress their own NACKs. In \cite{yamamoto2000performance}, Yamamoto et al. demonstrate that the delay incurred 
by a NACK-suppression mechanism does not affect the performance of NACK multicast control flow.

In 802.11 networks, selective NACKs can be used for the RTS/CTS handshake mechanism in order to reduces network 
congestion and packets collision. The result is a considerable throughput improvement and delay reduction 
\cite{impett2000receiver,sabah2010use}. In \cite{liu2005client}, NACKs at data-link layer are combined with 
NACKs at transport layer in order to improve video streaming performance over 3G cellular networks. In case 
of frame loss, a mobile device sends a selective data-link NACK to the base station. If the list frame has not 
been recovered after several successive NACKs, a transport-layer NACK is sent requesting resending the entire 
packet.

At transport layer, NACKs are used to provide reliable communications \cite{adamson2009nack,adamson2008multicast,
ichihara2003reliable,obraczka1998multicast}. \cite{adamson2009nack,adamson2008multicast} provide NACK-Oriented 
Reliable Multicast (NORM) Transport Protocol. NORM forms a reliable transport protocol between one or more 
senders to a group of receivers over an IP multicast network. In NORM, receivers use a selective NACK to 
notify senders about packets loss. A similar approach is used in \cite{ichihara2003reliable}, where NACKs are 
used as a packet loss detection mechanism in satellite communication. In this case, a NACK is generated by 
sending a signal. Senders detect NACK by monitoring the total electrical power in the frequency band used for 
uplink from the receiver. This kind of NACK enables several receivers to share a low-speed uplink circuit 
simultaneously preventing NACKs collision. In \cite{obraczka1998multicast}, Obraczka surveys multicast transport 
protocols summarizing NACK-based protocols, ACK-based protocols and some other hybrid approaches.

\section{Conclusions}
\label{sec:conclusion}
NDN and CCNx are two prominent ICN instances designed to address limitations of the current IP-based Internet. 
Network-layer NACKs are an important feature, adoption of which has been debated for both CCNx and NDN.
As we showed in this paper, NACKs can be beneficial in mitigating the impact of Interest Flooding attacks. 
Despite their benefits, we also showed that NACKs have certain challenging security implications. 
We identified two types of NACKs (\fnacks\ and \cnacks) and explored their security requirements. 
We then discussed how secure \cnacks\ can trigger producer-focused flooding attacks 
and discussed some potential methods for mitigating these attacks.

\balance

\bibliographystyle{IEEEtran}
\bibliography{IEEEabrv,references}

\end{document}